\documentclass[%
 reprint,
 groupedaddress,
 amsmath,amssymb,
 aps,
 pra,
 longbibliography
]{revtex4-2}

\usepackage[a4paper,margin=2.0cm]{geometry}
\usepackage{graphicx}
\usepackage{subfigure}
\usepackage{mathrsfs}
\usepackage{braket}
\usepackage{hyperref}
\usepackage{xcolor}
\usepackage{microtype}

\usepackage{pgfplots}
\pgfplotsset{compat=1.18}
\usepgfplotslibrary{groupplots,fillbetween}
\definecolor{tabblue}{RGB}{31,119,180}
\definecolor{taborange}{RGB}{255,127,14}
\definecolor{tabgreen}{RGB}{44,160,44}

\newcommand{\R}{\mathbb{R}}
\newcommand{\Tr}{\operatorname{Tr}}

\newcommand{\vect}[1]{\boldsymbol{#1}}

\begin{document}

\title{Shot-Based Quantum Encoding:\\
A Data-Loading Paradigm for Quantum Neural Networks}

\author{Basil Kyriacou}
\author{Viktoria Patapovich}
\author{Maniraman Periyasamy}
\author{Alexey Melnikov}\email[Corresponding author: ]{alexey@melnikov.info}

\affiliation{%
 Terra Quantum AG, 9000 St.~Gallen, Switzerland
}

% -------------------------- Abstract ----------------------------
\begin{abstract}
Efficient data loading remains a bottleneck for near-term quantum machine learning. Existing schemes (angle, amplitude, and basis encoding) either underuse the exponential Hilbert-space capacity or require circuit depths that exceed the coherence budgets of noisy intermediate-scale quantum hardware. We introduce shot-based quantum encoding (SBQE), a data embedding strategy that distributes the hardware's native resource, shots, according to a data-dependent classical distribution over multiple initial quantum states. By treating the shot counts as a learnable degree of freedom, SBQE produces a mixed-state representation whose expectation values are linear in the classical probabilities and can therefore be composed with nonlinear activation functions. We show that SBQE is structurally equivalent to a multilayer perceptron whose weights are realized by quantum circuits, and we describe a hardware-compatible implementation protocol. Benchmarks on three image datasets, with 10 independent initializations per model, show that SBQE achieves \(89.1\%\pm0.9\%\) test accuracy on Semeion (reducing error by \(5.3\%\) relative to amplitude encoding and matching a width-matched classical network), \(80.95\%\pm0.10\%\) on Fashion MNIST (exceeding amplitude encoding by \(+2.0\%\) and a linear multilayer perceptron by \(+1.3\%\)), and \(90.25\%\pm0.18\%\) on MNIST (exceeding amplitude encoding by 2.1 percentage points and the width-matched classical network by 0.3), all without any data-encoding gates.
\end{abstract}

\keywords{data encoding, mixed quantum states, NISQ, quantum machine learning, quantum neural networks, shot allocation, variational quantum circuits}

\maketitle

% ---- Published-version reference box (bottom of first page) ----
\begin{figure}[b!]
\noindent\fbox{\parbox{\dimexpr\columnwidth-2\fboxsep-2\fboxrule\relax}{\raggedright Please check the published version, which includes all the latest additions and corrections: Adv. Comput. 1:e70004, 2026, DOI: \href{https://doi.org/10.1002/adco.70004}{10.1002/adco.70004}}}
\end{figure}

% ================================================================
\section{Introduction}\label{sec:intro}
Quantum machine learning (QML) seeks to combine superposition, entanglement, and the exponential dimensionality of Hilbert space with classical learning algorithms \cite{Biamonte2017QuantumML,Cerezo2021VQA,qml_review_2023}.
Early results point to possible advantages in kernel evaluation \cite{Havlicek2019Supervised,Schuld2019Quantum}, generative modeling \cite{Benedetti_2019Generative,Zoufal_2019Quantum}, and combinatorial optimization and routing \cite{farhi2014qaoa,Egger_2021Warm,haboury2023routing}.
Yet two practical hurdles still limit what can be achieved with noisy-intermediate-scale quantum (NISQ) devices:
(i)~\emph{encoding} classical data efficiently into a quantum register, and
(ii)~\emph{extracting} task-relevant information through inherently noisy, sample-limited measurements.
The present work addresses the first challenge.

Current data-loading paradigms fall into two main families.
\emph{Angle encoding} assigns one feature to one qubit by mapping the feature value onto a single-qubit rotation \cite{peruzzo2014variational,schuld2021effect}.
This strategy benefits from shallow circuits and straightforward hardware implementation, but its representational capacity scales only linearly with the number of qubits, leaving the exponential Hilbert-space dimension largely untapped.
Data-reuploading techniques partially mitigate this limitation by inserting additional encoding layers throughout the variational circuit, at the cost of greater depth and noise susceptibility \cite{PerezSalinas2020DataReupload,Periyasamy_2022}.

\emph{Amplitude encoding}, in contrast, loads up to \(2^{n}\) classical amplitudes into an \(n\)-qubit state, thereby saturating the available quantum degrees of freedom \cite{Schuld2018Supervised}.
Unfortunately, exact state-preparation routines typically require \(\mathcal{O}(2^{n})\) controlled rotations and entangling gates, which exceed the coherence and fidelity budgets of current hardware \cite{Benedetti2019Parameterized}.

We introduce shot-based quantum encoding (SBQE), which bypasses the depth bottleneck by moving the encoding workload from coherent gates to the classical control layer that orchestrates repeated circuit executions (shots).
Instead of preparing a single pure state per datum, SBQE assigns each data point a classical probability vector that dictates how many shots start from each member of a fixed, hardware-friendly set of basis states.
The resulting statistical mixture constitutes the input to a shallow variational circuit; expectation values remain linear in the classical probabilities, so they compose directly with nonlinear classical post-processing and multilayer~QML architectures.
Because contemporary NISQ experiments already require thousands of repetitions to estimate observables with meaningful precision~\cite{abbas2020learn}, SBQE exploits an abundant resource without increasing coherent depth.

The remainder of this paper is organized as follows.
Section~\ref{sec:background} reviews competing data-loading strategies and situates SBQE within the broader QML literature.
Section~\ref{sec:SBQE} presents the density-matrix formulations of existing encodings, introduces the SBQE framework, and establishes its connection to multilayer perceptrons.
Numerical benchmarks on three image-classification tasks are presented in Section~\ref{sec:experiments}, and Sections~\ref{sec:discussion} and~\ref{sec:conclusion} discuss broader implications and limitations.

% ================================================================
\section{Background and Related Work}\label{sec:background}

Variational quantum circuits (VQCs) underpin most algorithms designed for the NISQ era.  A VQC applies a parameterized unitary \(U(\boldsymbol{\theta})\) to an initial state, measures a set of observables, and iteratively updates the angles \(\boldsymbol{\theta}\) to minimize a classical loss function \cite{peruzzo2014variational,Mitarai2018QCL,Cerezo2021VQA}.  Early exemplars such as the Variational Quantum Eigensolver and the Quantum Approximate Optimisation Algorithm \cite{farhi2014qaoa} validated the hybrid optimization loop, while subsequent studies analyzed gradient landscapes and identified the barren-plateau issue, namely an exponential decay of trainable signal with qubit count \cite{McClean2018Barren}.  Strategies including layer-wise training, problem-inspired ans\"atze, and noise-aware optimizers now aim to mitigate such pathologies.

The expressive power of any VQC ultimately depends on how classical information is embedded into the quantum register.  In \emph{angle encoding} a real-valued feature \(x\) modulates a single-qubit rotation, typically an \(R_{y}(x)\) or \(R_{z}(x)\) gate.  This linear mapping scales only with the number of qubits but enjoys shallow depth and straightforward hardware execution \cite{Havlicek2019Supervised,schuld2021effect}.  Depth-amplified variants, sometimes called ``quantum depth-infused'' (QDI) encodings, repeatedly interleave feature-dependent rotations with entangling layers to enrich the effective feature map \cite{PerezSalinas2020DataReupload}, and have been deployed in industrial prediction and control settings \cite{lee2025blastfurnace}.  By contrast, \emph{amplitude encoding} prepares a state whose computational-basis amplitudes store the entire data vector, achieving an exponential compression from \(2^{n}\) features into \(n\) qubits \cite{Schuld2018Supervised}.  Exact state preparation, however, requires gate counts that scale exponentially in the worst case; schemes based on uniformly controlled rotations \cite{mottonen2005transformation}, isometry decomposition \cite{Iten_2016isometries}, or quantum random access memory (QRAM) access \cite{Duan2024Compact} alleviate but do not eliminate the depth overhead.

An alternative line of research embraces \emph{probabilistic and mixed-state} descriptions.  Instead of loading information coherently, one prepares a classical or noisy ensemble of pure states whose density matrix encodes the data.  Such mixtures can bypass barren plateaus, exploit naturally occurring noise, or reduce coherent depth \cite{Larocca2022diagnosingbarren}.  Sampling-based protocols that randomize over easy-to-prepare states have been proposed for fair sampling and error mitigation \cite{Wang2021Noise}, while classical-shadow methods treat measurement randomness itself as a computational resource \cite{Huang2021Power}.  The present work builds on this perspective by turning the classical allocation of circuit repetitions (shots) into a learnable data-dependent degree of freedom, thereby unifying probabilistic encodings with the trainable expressivity of VQCs.

% ================================================================
\section{Shot-Based Quantum Encoding}\label{sec:SBQE}

\subsection{Encoding schemes in the density-matrix picture}

The output of a variational quantum circuit can be written in full generality as
\begin{equation}\label{eq:vqc_general}
    f(\vect{x},\vect{\theta})
    = \Tr\!\bigl[O\,\rho(\vect{x},\vect{\theta})\bigr],
\end{equation}
where $\rho$ is a density matrix and $O$ is a Hermitian observable.  Both $\rho$ and $O$ may depend on inputs $\vect{x}$ and trainable parameters $\vect{\theta}$ \cite{Schuld2018Supervised,nielsen2010quantum}.  The encoding strategy determines how $\rho$ is constructed from a reference state $\ket{0}$.

\emph{Basis encoding} maps an integer $i\in\{0,\dots,2^{n}-1\}$ to the corresponding computational-basis state by flipping individual qubits with Pauli-$X$ gates:
\begin{equation}
    \rho(i,\vect{\theta})
    = U(\vect{\theta})\,\ket{i}\!\bra{i}\,U^{\dagger}(\vect{\theta}),
    \quad i\in\mathbb{N}.
\end{equation}

\emph{Angle encoding} embeds continuous features through single-qubit rotations, making the unitary itself data-dependent.  Because the rotations act at the individual-qubit level, the number of encodable features scales linearly with the number of qubits, in contrast to basis, amplitude, and shot-based encoding, all of which can accommodate exponentially more features than qubits~\cite{schuld2021effect,kordzanganeh2023exponentially}:
\begin{equation}
    \rho(\vect{x},\vect{\theta})
    = U(\vect{x},\vect{\theta})\,\ket{0}\!\bra{0}\,
      U^{\dagger}(\vect{x},\vect{\theta}).
\end{equation}

\emph{Amplitude encoding} places an $L^{2}$-normalized data vector directly into the probability amplitudes of the state, compressing up to $2^{n}$ features into $n$ qubits at the cost of an exponentially deep preparation circuit \cite{Schuld2018Supervised}:
\begin{equation}
    \rho(\vect{x},\vect{\theta})
    = U(\vect{\theta})\,\ket{\vect{x}}\!\bra{\vect{x}}\,
      U^{\dagger}(\vect{\theta}).
\end{equation}

All three schemes above produce \emph{pure} states.  SBQE departs from this pattern by encoding data into a \emph{mixed} state whose mixture weights are themselves the data representation.

\subsection{Density-matrix formulation of SBQE}

Let $\{\ket{\psi_j}\}_{j=1}^n$ be a fixed, efficiently preparable set of pure states on $q$ qubits.
SBQE maps a datum $\vect{x}$ to a probability vector
${\vect{p}(\vect{x})\!=\!(p_1,\dots,p_n)}$ with $\sum_j p_j=1$.
Executing $N_\text{tot}$ shots, we draw integer counts
$\vect{N}(\vect{x}) = (N_1,\dots,N_n)$ from $\text{Multinomial}(N_\text{tot},\vect{p}(\vect{x}))$
and initialize exactly $N_j$ runs in $\ket{\psi_j}$.
The encoded state is therefore the classical mixture
\begin{equation}
    \rho(\vect{x}) \;=\; \sum_{j=1}^{n} p_j(\vect{x}) \,
    \ket{\psi_j}\!\bra{\psi_j}.
    \label{eq:rho_SBQE}
\end{equation}
Subsequent application of a variational unitary $U(\vect{\theta})$ and measurement of Hermitian observables $\{O_i\}_{i=1}^{m}$ yields outputs
\begin{equation}
    f_i(\vect{x},\vect{\theta}) \;=\;
    \Tr\!\bigl[ O_i U(\vect{\theta}) \rho(\vect{x}) U^{\dagger}(\vect{\theta}) \bigr].
    \label{eq:SBQE_forward}
\end{equation}

As a concrete example, consider a single-qubit register with two initial states $\ket{0}\!\bra{0}$ and $\ket{1}\!\bra{1}$ and a total budget of $N_\text{tot}=1024$ shots.  For a datum~$\vect{x}$ whose probability vector is $\vect{p}(\vect{x})=(2/3,\;1/3)$, the experiment allocates $\vect{N}=(683,\,341)$ shots to $\ket{0}\!\bra{0}$ and $\ket{1}\!\bra{1}$, respectively.
If the initial-state pool is enlarged to $\{\ket{0}\!\bra{0},\,\ket{1}\!\bra{1},\,\ket{+}\!\bra{+},\,\ket{-}\!\bra{-}\}$, the probability vector has four components, for example, $\vect{p}=(3/8,\,1/4,\,1/4,\,1/8)$, giving $\vect{N}=(384,\,256,\,256,\,128)$.
This contrasts with the standard practice of starting all 1024 shots in~$\ket{0}\!\bra{0}$; here the data are encoded entirely through the classical redistribution of shots, without any additional quantum gates.

When the initial states are the $2^{q}$ computational-basis states and the observables are projective measurements in the same basis, Eq.~\eqref{eq:SBQE_forward} simplifies to $f_i = |U_{ij}(\vect{\theta})|^{2} p_j(\vect{x})$, so that~$\vect{f}$ is itself a probability vector that can be fed directly into a subsequent SBQE layer.

\begin{figure*}[t]
  \centering
  \includegraphics[width=\linewidth]{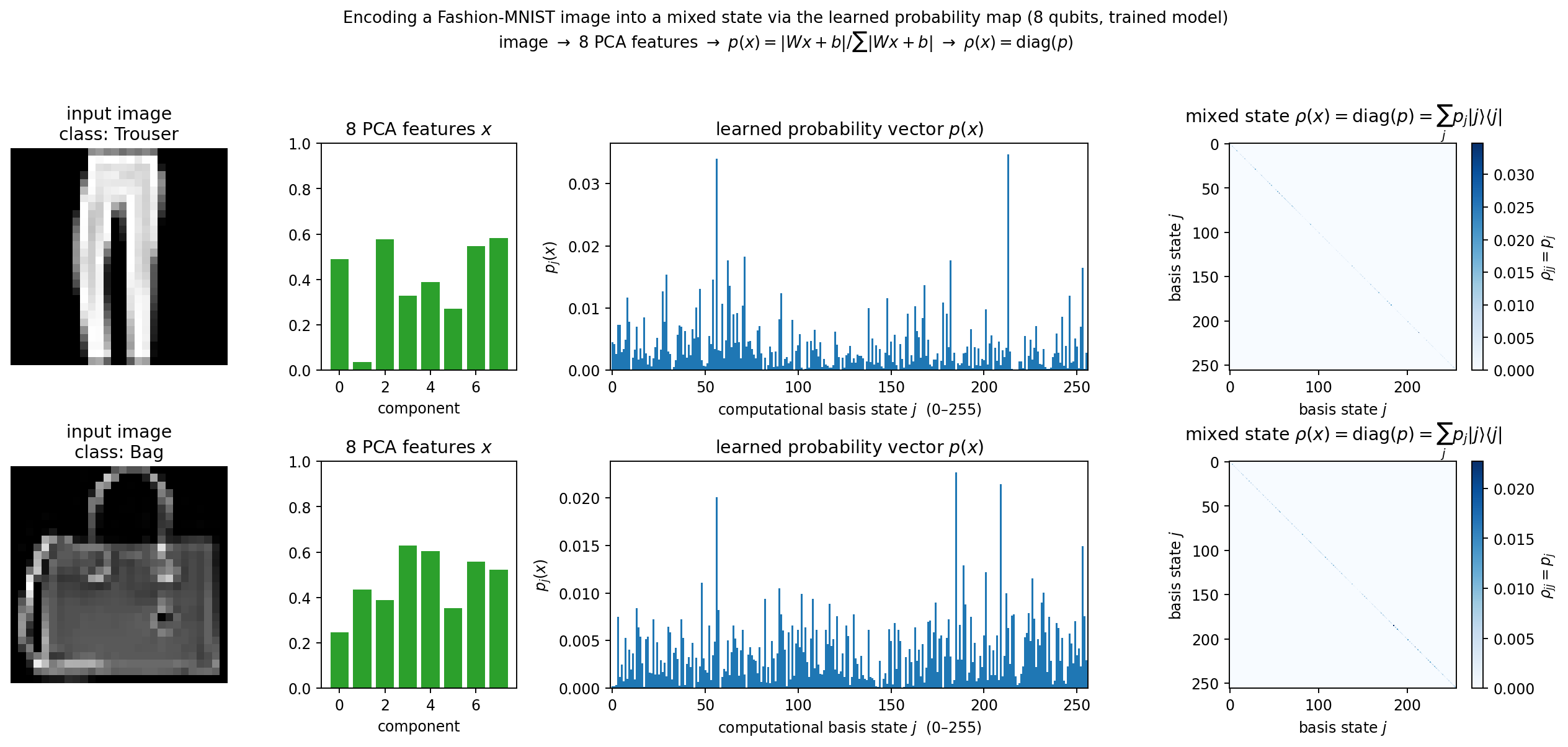}
  \caption{Encoding a datum into a mixed state through the learned probability map ($q=8$ qubits, trained model).  For two Fashion-MNIST images of different classes (a trouser and a bag), from left to right: the input image; its $d=8$ principal-component features $\vect{x}$; the learned probability vector $\vect{p}(\vect{x})$ over the $2^{8}=256$ computational-basis states [Eq.~\eqref{eq:simplex}]; and the induced mixed state $\rho(\vect{x})=\operatorname{diag}(\vect{p}(\vect{x}))=\sum_j p_j\ket{j}\!\bra{j}$, which is diagonal in the computational basis.  Images of different classes yield distinct probability vectors, and hence distinct mixed states.}
  \label{fig:px_example}
\end{figure*}

In the experiments reported below, the probability vector $\vect{p}(\vect{x})$ is not fixed in advance but is produced by a trainable map applied to the reduced data, so that the encoding is adapted to the task rather than prescribed by a fixed pixel-to-probability rule.  Each datum is first reduced to its $d=8$ leading principal components $\vect{x}\in\R^{d}$ (Section~\ref{sec:experiments}) and rescaled to the unit interval; a trainable affine layer maps it to $\vect{z}=W\vect{x}+\vect{b}\in\R^{n}$, and an element-wise absolute value followed by $\ell_1$ normalization projects $\vect{z}$ onto the probability simplex,
\begin{equation}
    p_j(\vect{x}) \;=\; \frac{|z_j|}{\sum_k |z_k|},
    \qquad p_j \ge 0, \quad \sum_j p_j = 1.
    \label{eq:simplex}
\end{equation}
The weights $W$ and biases $\vect{b}$ are optimized jointly with the circuit parameters $\vect{\theta}$ by back-propagation.  Taking the pool $\{\ket{\psi_j}\}$ to be the full computational basis, so that $n=2^{q}$, the entries of $\vect{p}(\vect{x})$ are the populations of the $2^{q}$ basis states and the encoded state is diagonal in that basis, $\rho(\vect{x})=\sum_j p_j(\vect{x})\ket{j}\!\bra{j}=\operatorname{diag}(\vect{p}(\vect{x}))$; it is realized physically by allocating a fraction $p_j$ of the $N_\text{tot}$ shots to $\ket{j}$ (prepared with single-qubit $X$ gates), with no data-dependent rotation or entangling gates.  Figure~\ref{fig:px_example} illustrates this pipeline on two trained Fashion-MNIST examples.

\subsection{Connection to multilayer perceptrons}
Inserting Eq.~\eqref{eq:rho_SBQE} into Eq.~\eqref{eq:SBQE_forward},
\begin{equation*}
    f_i(\vect{x},\vect{\theta})
    = \sum_{j=1}^{n} W_{ij}(\vect{\theta}) \, p_j(\vect{x})
    \label{eq:fi_def}
\end{equation*}

\begin{equation}
    W_{ij}(\vect{\theta})
    := \bra{\psi_j} U^{\dagger}(\vect{\theta}) O_i U(\vect{\theta}) \ket{\psi_j}
    \label{eq:Wij_def}
\end{equation}

Hence each layer is a \emph{linear} map in the encoded probabilities, analogous to the affine transformation in classical MLPs \cite{rosenblatt1958perceptron}.
Placing the classical and quantum cases side by side makes the correspondence explicit:
\begin{align}
    \text{Classical:}\quad & n_i = \sigma\!\bigl(\widetilde{W}_{ij}(\vect{\theta})\,x_j\bigr), \\
    \text{Quantum:}\quad  & n_i = \sigma\!\bigl(W_{ij}(\vect{\theta})\,p_j(\vect{x})\bigr).
\end{align}
The key differences are that in the quantum case the data must first be processed into a probability vector $\vect{p}(\vect{x})$, and that $\widetilde{W}_{ij}$ stores trainable weights directly, whereas $W_{ij}$ is the output of a quantum device and depends on $\vect{\theta}$ only indirectly through $U(\vect{\theta})$.

A nonlinearity $\sigma$ that maps its input back to a valid probability vector allows the output of one layer to serve as the shot distribution for the next, giving a multilayer architecture:
\begin{equation}
    \vect{p}^{(k+1)}
    = \sigma\!\bigl( W^{(k)}(\vect{\theta}^{(k)}) \, \vect{p}^{(k)} \bigr),
\end{equation}
with $\vect{p}^{(0)}\equiv \vect{p}(\vect{x})$.
SBQE therefore supports stacked ``quantum perceptrons'' in which \emph{all} trainable weights reside in quantum parameters $\vect{\theta}$.

One such activation function that preserves the probability-vector structure is the \emph{log-ReLU cascade}:
\begin{align}
    \vect{p}' &= \log \vect{p}, \notag\\
    \vect{p}'' &= \vect{p}' - \langle \vect{p}' \rangle, \notag\\
    \vect{p}''' &= \operatorname{ReLU}(\vect{p}''), \notag\\
    \sigma(\vect{p}) &= \vect{p}'''/\textstyle\sum_j p'''_{j}.
    \label{eq:log_relu}
\end{align}
Two edge cases require special treatment: if $p_i=0$, the entry is set to zero and excluded from the remaining steps; if $\vect{p}'''=\vect{0}$, the output defaults to a uniform distribution over the surviving entries.  This function zeros out low-probability components, which is well suited for finite-shot experiments where small probabilities cannot be resolved anyway.

%================================================================
\section{Numerical Experiments}\label{sec:experiments}

\subsection{Benchmarks}

We evaluate SBQE in a machine learning setting on three widely used image-classification suites that differ in visual complexity and class balance.
\emph{Fashion-MNIST} offers 70\,000 $28\times28$ grayscale product images drawn from 10 garment categories \cite{xiao2017fashionmnist}.
\emph{Semeion} comprises 1593 handwritten digits collected from roughly 80 individuals and pre-binarized to 256 input pixels \cite{semeion_handwritten_digit_178}.
\emph{MNIST} comprises 70\,000 $28\times28$ grayscale images of handwritten digits \cite{lecun1998gradient}.
All three tasks are cast as 10-class classification.

Each raw image is vectorized, centered, and projected onto the first $d=8$ principal components via PCA, reducing the dimensionality to match an eight-qubit register.
Principal-component analysis~\cite{Pearson1901PCA} is implemented with {\tt scikit-learn}'s exact SVD routine (double precision).
The same $d=8$ reduction and eight-qubit register are used for MNIST as for the other datasets; this is distinct from the illustrative 784-pixel, 10-qubit configuration mentioned in Section~\ref{sec:discussion}.

\subsection{Experimental Setup}

\paragraph*{Model zoo.}
We compare three architectures that share the same quantum workload (\(n_{\mathrm{q}}=8\) qubits, \(n_{\mathrm{layers}}=4\)) and are matched for \emph{total trainable parameters} wherever applicable.

\begin{enumerate}
  \item \emph{Amplitude Hybrid}: baseline with deterministic amplitude embedding followed by variational layers with $\mathrm{Rot}$-CNOT ladder, identical to the template in \cite{bergholm2022pennylane}.
  \item \emph{SBQE Hybrid}: our SBQE variant: an input dense layer produces a simplex-normalized probability vector, which is fed to the same variational circuit as above.
  \item \emph{Width-Matched Linear}: a two-layer classical MLP whose hidden width is computed analytically so that its parameter count matches that of the quantum hybrids as closely as possible without exceeding it.
\end{enumerate}

\paragraph*{Circuit details.}
All quantum layers employ the $\mathrm{Rot}(\alpha,\beta,\gamma)$ gate per qubit followed by linear-chain CNOTs; parameters are initialized from~$\mathcal{N}(0,0.02)$.
Statevector back-propagation is provided by {\tt Pennylane~0.41}~\cite{bergholm2022pennylane} and executed on a single NVIDIA A100 GPU via {\tt JAX~0.4.28}~\cite{jax2018github}; comparable experiments can be reproduced with the TQml simulator~\cite{kuzmin2025tqml}.
Classical submodules are written in {\tt Flax~0.8}.

\paragraph*{Optimization and regularization.}
All models minimize the cross-entropy loss with the Lion optimizer (\(\beta_1=0.95,\beta_2=0.98\), decoupled weight decay \(10^{-2}\); batch size 7000) over 20\,000 epochs.
Gradients are clipped to a global norm of 0.01.

\paragraph*{Training protocol.}
For each dataset, we draw stratified splits of 70\%/15\%/15\% for training, validation, and testing, fixing the random seed across baselines.
Every configuration is run 10 times with independent parameter initialization, and metrics are averaged.
Early stopping is \emph{disabled} in order to expose asymptotic behavior; instead we report the best validation-epoch weights on the held-out test set.

\begin{figure*}[t]
  \centering
  % Auto-generated (Option B, pgfplots): learning curves for Semeion.
% Preamble requirements are listed in figs_tex/README_pgfplots.txt.
\resizebox{\linewidth}{!}{%
\begin{tikzpicture}
\begin{groupplot}[
    group style={group size=2 by 1, horizontal sep=1.6cm},
    width=7cm, height=5.2cm,
    xmode=log, ymode=log,
    xlabel={Epoch},
    grid=both, grid style={gray!25},
    tick align=outside, tick pos=left,
    label style={font=\small}, tick label style={font=\footnotesize},
]
\nextgroupplot[ylabel={Loss}]
\addplot[draw=none,forget plot,name path=semeion0tlhi] table[header=false,x index=0,y index=3]{figs_tex/data_semeion.dat};
\addplot[draw=none,forget plot,name path=semeion0tllo] table[header=false,x index=0,y index=2]{figs_tex/data_semeion.dat};
\addplot[tabblue,opacity=0.15,forget plot] fill between[of=semeion0tlhi and semeion0tllo];
\addplot[draw=none,forget plot,name path=semeion0vlhi] table[header=false,x index=0,y index=6]{figs_tex/data_semeion.dat};
\addplot[draw=none,forget plot,name path=semeion0vllo] table[header=false,x index=0,y index=5]{figs_tex/data_semeion.dat};
\addplot[tabblue,opacity=0.15,forget plot] fill between[of=semeion0vlhi and semeion0vllo];
\addplot[draw=none,forget plot,name path=semeion1tlhi] table[header=false,x index=0,y index=15]{figs_tex/data_semeion.dat};
\addplot[draw=none,forget plot,name path=semeion1tllo] table[header=false,x index=0,y index=14]{figs_tex/data_semeion.dat};
\addplot[taborange,opacity=0.15,forget plot] fill between[of=semeion1tlhi and semeion1tllo];
\addplot[draw=none,forget plot,name path=semeion1vlhi] table[header=false,x index=0,y index=18]{figs_tex/data_semeion.dat};
\addplot[draw=none,forget plot,name path=semeion1vllo] table[header=false,x index=0,y index=17]{figs_tex/data_semeion.dat};
\addplot[taborange,opacity=0.15,forget plot] fill between[of=semeion1vlhi and semeion1vllo];
\addplot[draw=none,forget plot,name path=semeion2tlhi] table[header=false,x index=0,y index=27]{figs_tex/data_semeion.dat};
\addplot[draw=none,forget plot,name path=semeion2tllo] table[header=false,x index=0,y index=26]{figs_tex/data_semeion.dat};
\addplot[tabgreen,opacity=0.15,forget plot] fill between[of=semeion2tlhi and semeion2tllo];
\addplot[draw=none,forget plot,name path=semeion2vlhi] table[header=false,x index=0,y index=30]{figs_tex/data_semeion.dat};
\addplot[draw=none,forget plot,name path=semeion2vllo] table[header=false,x index=0,y index=29]{figs_tex/data_semeion.dat};
\addplot[tabgreen,opacity=0.15,forget plot] fill between[of=semeion2vlhi and semeion2vllo];
\addplot[tabblue,solid,thick,forget plot] table[header=false,x index=0,y index=1]{figs_tex/data_semeion.dat};
\addplot[tabblue,dashed,thick,forget plot] table[header=false,x index=0,y index=4]{figs_tex/data_semeion.dat};
\addplot[taborange,solid,thick,forget plot] table[header=false,x index=0,y index=13]{figs_tex/data_semeion.dat};
\addplot[taborange,dashed,thick,forget plot] table[header=false,x index=0,y index=16]{figs_tex/data_semeion.dat};
\addplot[tabgreen,solid,thick,forget plot] table[header=false,x index=0,y index=25]{figs_tex/data_semeion.dat};
\addplot[tabgreen,dashed,thick,forget plot] table[header=false,x index=0,y index=28]{figs_tex/data_semeion.dat};
\nextgroupplot[ylabel={Error (1 - Accuracy)}, ymin=0.0005]
\addplot[draw=none,forget plot,name path=semeion0tehi] table[header=false,x index=0,y index=9]{figs_tex/data_semeion.dat};
\addplot[draw=none,forget plot,name path=semeion0telo] table[header=false,x index=0,y index=8]{figs_tex/data_semeion.dat};
\addplot[tabblue,opacity=0.15,forget plot] fill between[of=semeion0tehi and semeion0telo];
\addplot[draw=none,forget plot,name path=semeion0vehi] table[header=false,x index=0,y index=12]{figs_tex/data_semeion.dat};
\addplot[draw=none,forget plot,name path=semeion0velo] table[header=false,x index=0,y index=11]{figs_tex/data_semeion.dat};
\addplot[tabblue,opacity=0.15,forget plot] fill between[of=semeion0vehi and semeion0velo];
\addplot[draw=none,forget plot,name path=semeion1tehi] table[header=false,x index=0,y index=21]{figs_tex/data_semeion.dat};
\addplot[draw=none,forget plot,name path=semeion1telo] table[header=false,x index=0,y index=20]{figs_tex/data_semeion.dat};
\addplot[taborange,opacity=0.15,forget plot] fill between[of=semeion1tehi and semeion1telo];
\addplot[draw=none,forget plot,name path=semeion1vehi] table[header=false,x index=0,y index=24]{figs_tex/data_semeion.dat};
\addplot[draw=none,forget plot,name path=semeion1velo] table[header=false,x index=0,y index=23]{figs_tex/data_semeion.dat};
\addplot[taborange,opacity=0.15,forget plot] fill between[of=semeion1vehi and semeion1velo];
\addplot[draw=none,forget plot,name path=semeion2tehi] table[header=false,x index=0,y index=33]{figs_tex/data_semeion.dat};
\addplot[draw=none,forget plot,name path=semeion2telo] table[header=false,x index=0,y index=32]{figs_tex/data_semeion.dat};
\addplot[tabgreen,opacity=0.15,forget plot] fill between[of=semeion2tehi and semeion2telo];
\addplot[draw=none,forget plot,name path=semeion2vehi] table[header=false,x index=0,y index=36]{figs_tex/data_semeion.dat};
\addplot[draw=none,forget plot,name path=semeion2velo] table[header=false,x index=0,y index=35]{figs_tex/data_semeion.dat};
\addplot[tabgreen,opacity=0.15,forget plot] fill between[of=semeion2vehi and semeion2velo];
\addplot[tabblue,solid,thick,forget plot] table[header=false,x index=0,y index=7]{figs_tex/data_semeion.dat};
\addplot[tabblue,dashed,thick,forget plot] table[header=false,x index=0,y index=10]{figs_tex/data_semeion.dat};
\addplot[taborange,solid,thick,forget plot] table[header=false,x index=0,y index=19]{figs_tex/data_semeion.dat};
\addplot[taborange,dashed,thick,forget plot] table[header=false,x index=0,y index=22]{figs_tex/data_semeion.dat};
\addplot[tabgreen,solid,thick,forget plot] table[header=false,x index=0,y index=31]{figs_tex/data_semeion.dat};
\addplot[tabgreen,dashed,thick,forget plot] table[header=false,x index=0,y index=34]{figs_tex/data_semeion.dat};
\end{groupplot}
% Shared legend, both rows centred on the two-panel midline (row 1: three colours; row 2: line styles).
\path (group c1r1.north west) -- (group c2r1.north east) coordinate[midway] (legsemeionmid);
\node[anchor=south, font=\small] at ([yshift=4mm]legsemeionmid) {%
  \begin{tabular}{@{}c@{}}
    \tikz\draw[tabblue,thick](0,0)--(0.45,0);~SBQE\hspace{1.2em}%
    \tikz\draw[taborange,thick](0,0)--(0.45,0);~Ampl. encoding\hspace{1.2em}%
    \tikz\draw[tabgreen,thick](0,0)--(0.45,0);~MLP\\[2pt]
    \tikz\draw[black,thick](0,0)--(0.45,0);~train\hspace{1.2em}%
    \tikz\draw[black,thick,dashed](0,0)--(0.45,0);~validation
  \end{tabular}};
\end{tikzpicture}%
}
  \caption{Learning curves on the \textsc{Semeion} benchmark ($8$ qubits, $4$ layers).
  Left panel: cross-entropy loss; right panel: classification error ($1-$accuracy).
  Within each panel, solid lines are training and dashed lines validation, both averaged over 10 runs; shaded bands show one standard deviation, and colors distinguish the three models as in the legend.}
  \label{fig:semeion_curves}
\end{figure*}

\begin{figure*}[t]
  \centering
  % Auto-generated (Option B, pgfplots): learning curves for Fashion-MNIST.
% Preamble requirements are listed in figs_tex/README_pgfplots.txt.
\resizebox{\linewidth}{!}{%
\begin{tikzpicture}
\begin{groupplot}[
    group style={group size=2 by 1, horizontal sep=1.6cm},
    width=7cm, height=5.2cm,
    xmode=log, ymode=log,
    xlabel={Epoch},
    grid=both, grid style={gray!25},
    tick align=outside, tick pos=left,
    yticklabel={\pgfmathparse{exp(\tick)}\pgfmathprintnumber[fixed,precision=2]{\pgfmathresult}},
    label style={font=\small}, tick label style={font=\footnotesize},
]
\nextgroupplot[ylabel={Loss}]
\addplot[draw=none,forget plot,name path=fashion0tlhi] table[header=false,x index=0,y index=3]{figs_tex/data_fashion.dat};
\addplot[draw=none,forget plot,name path=fashion0tllo] table[header=false,x index=0,y index=2]{figs_tex/data_fashion.dat};
\addplot[tabblue,opacity=0.15,forget plot] fill between[of=fashion0tlhi and fashion0tllo];
\addplot[draw=none,forget plot,name path=fashion0vlhi] table[header=false,x index=0,y index=6]{figs_tex/data_fashion.dat};
\addplot[draw=none,forget plot,name path=fashion0vllo] table[header=false,x index=0,y index=5]{figs_tex/data_fashion.dat};
\addplot[tabblue,opacity=0.15,forget plot] fill between[of=fashion0vlhi and fashion0vllo];
\addplot[draw=none,forget plot,name path=fashion1tlhi] table[header=false,x index=0,y index=15]{figs_tex/data_fashion.dat};
\addplot[draw=none,forget plot,name path=fashion1tllo] table[header=false,x index=0,y index=14]{figs_tex/data_fashion.dat};
\addplot[taborange,opacity=0.15,forget plot] fill between[of=fashion1tlhi and fashion1tllo];
\addplot[draw=none,forget plot,name path=fashion1vlhi] table[header=false,x index=0,y index=18]{figs_tex/data_fashion.dat};
\addplot[draw=none,forget plot,name path=fashion1vllo] table[header=false,x index=0,y index=17]{figs_tex/data_fashion.dat};
\addplot[taborange,opacity=0.15,forget plot] fill between[of=fashion1vlhi and fashion1vllo];
\addplot[draw=none,forget plot,name path=fashion2tlhi] table[header=false,x index=0,y index=27]{figs_tex/data_fashion.dat};
\addplot[draw=none,forget plot,name path=fashion2tllo] table[header=false,x index=0,y index=26]{figs_tex/data_fashion.dat};
\addplot[tabgreen,opacity=0.15,forget plot] fill between[of=fashion2tlhi and fashion2tllo];
\addplot[draw=none,forget plot,name path=fashion2vlhi] table[header=false,x index=0,y index=30]{figs_tex/data_fashion.dat};
\addplot[draw=none,forget plot,name path=fashion2vllo] table[header=false,x index=0,y index=29]{figs_tex/data_fashion.dat};
\addplot[tabgreen,opacity=0.15,forget plot] fill between[of=fashion2vlhi and fashion2vllo];
\addplot[tabblue,solid,thick,forget plot] table[header=false,x index=0,y index=1]{figs_tex/data_fashion.dat};
\addplot[tabblue,dashed,thick,forget plot] table[header=false,x index=0,y index=4]{figs_tex/data_fashion.dat};
\addplot[taborange,solid,thick,forget plot] table[header=false,x index=0,y index=13]{figs_tex/data_fashion.dat};
\addplot[taborange,dashed,thick,forget plot] table[header=false,x index=0,y index=16]{figs_tex/data_fashion.dat};
\addplot[tabgreen,solid,thick,forget plot] table[header=false,x index=0,y index=25]{figs_tex/data_fashion.dat};
\addplot[tabgreen,dashed,thick,forget plot] table[header=false,x index=0,y index=28]{figs_tex/data_fashion.dat};
\nextgroupplot[ylabel={Error (1 - Accuracy)}]
\addplot[draw=none,forget plot,name path=fashion0tehi] table[header=false,x index=0,y index=9]{figs_tex/data_fashion.dat};
\addplot[draw=none,forget plot,name path=fashion0telo] table[header=false,x index=0,y index=8]{figs_tex/data_fashion.dat};
\addplot[tabblue,opacity=0.15,forget plot] fill between[of=fashion0tehi and fashion0telo];
\addplot[draw=none,forget plot,name path=fashion0vehi] table[header=false,x index=0,y index=12]{figs_tex/data_fashion.dat};
\addplot[draw=none,forget plot,name path=fashion0velo] table[header=false,x index=0,y index=11]{figs_tex/data_fashion.dat};
\addplot[tabblue,opacity=0.15,forget plot] fill between[of=fashion0vehi and fashion0velo];
\addplot[draw=none,forget plot,name path=fashion1tehi] table[header=false,x index=0,y index=21]{figs_tex/data_fashion.dat};
\addplot[draw=none,forget plot,name path=fashion1telo] table[header=false,x index=0,y index=20]{figs_tex/data_fashion.dat};
\addplot[taborange,opacity=0.15,forget plot] fill between[of=fashion1tehi and fashion1telo];
\addplot[draw=none,forget plot,name path=fashion1vehi] table[header=false,x index=0,y index=24]{figs_tex/data_fashion.dat};
\addplot[draw=none,forget plot,name path=fashion1velo] table[header=false,x index=0,y index=23]{figs_tex/data_fashion.dat};
\addplot[taborange,opacity=0.15,forget plot] fill between[of=fashion1vehi and fashion1velo];
\addplot[draw=none,forget plot,name path=fashion2tehi] table[header=false,x index=0,y index=33]{figs_tex/data_fashion.dat};
\addplot[draw=none,forget plot,name path=fashion2telo] table[header=false,x index=0,y index=32]{figs_tex/data_fashion.dat};
\addplot[tabgreen,opacity=0.15,forget plot] fill between[of=fashion2tehi and fashion2telo];
\addplot[draw=none,forget plot,name path=fashion2vehi] table[header=false,x index=0,y index=36]{figs_tex/data_fashion.dat};
\addplot[draw=none,forget plot,name path=fashion2velo] table[header=false,x index=0,y index=35]{figs_tex/data_fashion.dat};
\addplot[tabgreen,opacity=0.15,forget plot] fill between[of=fashion2vehi and fashion2velo];
\addplot[tabblue,solid,thick,forget plot] table[header=false,x index=0,y index=7]{figs_tex/data_fashion.dat};
\addplot[tabblue,dashed,thick,forget plot] table[header=false,x index=0,y index=10]{figs_tex/data_fashion.dat};
\addplot[taborange,solid,thick,forget plot] table[header=false,x index=0,y index=19]{figs_tex/data_fashion.dat};
\addplot[taborange,dashed,thick,forget plot] table[header=false,x index=0,y index=22]{figs_tex/data_fashion.dat};
\addplot[tabgreen,solid,thick,forget plot] table[header=false,x index=0,y index=31]{figs_tex/data_fashion.dat};
\addplot[tabgreen,dashed,thick,forget plot] table[header=false,x index=0,y index=34]{figs_tex/data_fashion.dat};
\end{groupplot}
% Shared legend, both rows centred on the two-panel midline (row 1: three colours; row 2: line styles).
\path (group c1r1.north west) -- (group c2r1.north east) coordinate[midway] (legfashionmid);
\node[anchor=south, font=\small] at ([yshift=4mm]legfashionmid) {%
  \begin{tabular}{@{}c@{}}
    \tikz\draw[tabblue,thick](0,0)--(0.45,0);~SBQE\hspace{1.2em}%
    \tikz\draw[taborange,thick](0,0)--(0.45,0);~Ampl. encoding\hspace{1.2em}%
    \tikz\draw[tabgreen,thick](0,0)--(0.45,0);~MLP\\[2pt]
    \tikz\draw[black,thick](0,0)--(0.45,0);~train\hspace{1.2em}%
    \tikz\draw[black,thick,dashed](0,0)--(0.45,0);~validation
  \end{tabular}};
\end{tikzpicture}%
}
  \caption{Learning curves on the \textsc{Fashion-MNIST} benchmark ($8$ qubits, $4$ layers).
  Left panel: cross-entropy loss; right panel: classification error ($1-$accuracy).
  Within each panel, solid lines are training and dashed lines validation, both averaged over 10 runs; shaded bands show one standard deviation, and colors distinguish the three models as in the legend.}
  \label{fig:fashion_curves}
\end{figure*}

% ================================================================
\subsection{Results}\label{sec:results}

We trained each architecture described in Section~\ref{sec:experiments} for 10 independent random initializations and report test-set performance as mean$\,\pm\,$standard deviation.  Convergence trajectories are shown in Figure~\ref{fig:semeion_curves} (Semeion), Figure~\ref{fig:fashion_curves} (Fashion-MNIST), and Figure~\ref{fig:mnist_curves} (MNIST).

\paragraph*{Semeion handwritten digits.}
SBQE attains a test accuracy of
\(89.1\%\pm0.9\%\) and a cross-entropy loss of
\(0.428\pm0.037\).  This represents a relative error reduction of
\(5.3\%\) against amplitude encoding at identical depth and qubit count,
and narrows the gap to a width-matched classical MLP
(\(89.6\%\pm1.3\%\)) to within statistical noise.
Although the state-of-the-art CNNs exceed \(95\%\) on this dataset
\cite{Satbhaya2025SemeionCNN}, such models employ
\(>10^{5}\) parameters and convolutional inductive bias, whereas all three
baselines here are capped below \(1.1\times10^{4}\) parameters.
The result therefore demonstrates that reallocating
shot resources already available on NISQ hardware can recover essentially all
of the performance given by purely classical depth.

\paragraph*{Fashion-MNIST.}
On the harder Fashion benchmark, SBQE again leads with
\(80.95\%\pm0.10\%\) accuracy, outperforming amplitude encoding by an
absolute \(2.0\%\) and the linear MLP by \(1.3\%\).
On the same eight PCA features a purely linear classifier (multinomial
logistic regression) reaches only \(73.5\%\), so SBQE adds 7.5 percentage
points on identical inputs, whereas amplitude-encoded quantum classifiers
reported in the literature rely on deeper data-reuploading circuits and
achieve \(78\%\)--\(79\%\) at eight qubits
\cite{shen2024fashionmnist,senokosov2024quantum}.  A linear classifier
reaches \(85\%\) only when given the full \(784\)-pixel input rather than
the eight components used here \cite{Kaggle025FashionMNIST}.

\begin{figure*}[t]
  \centering
  % Auto-generated (Option B, pgfplots): learning curves for MNIST.
% Preamble requirements are listed in figs_tex/README_pgfplots.txt.
\resizebox{\linewidth}{!}{%
\begin{tikzpicture}
\begin{groupplot}[
    group style={group size=2 by 1, horizontal sep=1.6cm},
    width=7cm, height=5.2cm,
    xmode=log, ymode=log,
    xlabel={Epoch},
    grid=both, grid style={gray!25},
    tick align=outside, tick pos=left,
    yticklabel={\pgfmathparse{exp(\tick)}\pgfmathprintnumber[fixed,precision=2]{\pgfmathresult}},
    label style={font=\small}, tick label style={font=\footnotesize},
]
\nextgroupplot[ylabel={Loss}]
\addplot[draw=none,forget plot,name path=mnist0tlhi] table[header=false,x index=0,y index=3]{figs_tex/data_mnist.dat};
\addplot[draw=none,forget plot,name path=mnist0tllo] table[header=false,x index=0,y index=2]{figs_tex/data_mnist.dat};
\addplot[tabblue,opacity=0.15,forget plot] fill between[of=mnist0tlhi and mnist0tllo];
\addplot[draw=none,forget plot,name path=mnist0vlhi] table[header=false,x index=0,y index=6]{figs_tex/data_mnist.dat};
\addplot[draw=none,forget plot,name path=mnist0vllo] table[header=false,x index=0,y index=5]{figs_tex/data_mnist.dat};
\addplot[tabblue,opacity=0.15,forget plot] fill between[of=mnist0vlhi and mnist0vllo];
\addplot[draw=none,forget plot,name path=mnist1tlhi] table[header=false,x index=0,y index=15]{figs_tex/data_mnist.dat};
\addplot[draw=none,forget plot,name path=mnist1tllo] table[header=false,x index=0,y index=14]{figs_tex/data_mnist.dat};
\addplot[taborange,opacity=0.15,forget plot] fill between[of=mnist1tlhi and mnist1tllo];
\addplot[draw=none,forget plot,name path=mnist1vlhi] table[header=false,x index=0,y index=18]{figs_tex/data_mnist.dat};
\addplot[draw=none,forget plot,name path=mnist1vllo] table[header=false,x index=0,y index=17]{figs_tex/data_mnist.dat};
\addplot[taborange,opacity=0.15,forget plot] fill between[of=mnist1vlhi and mnist1vllo];
\addplot[draw=none,forget plot,name path=mnist2tlhi] table[header=false,x index=0,y index=27]{figs_tex/data_mnist.dat};
\addplot[draw=none,forget plot,name path=mnist2tllo] table[header=false,x index=0,y index=26]{figs_tex/data_mnist.dat};
\addplot[tabgreen,opacity=0.15,forget plot] fill between[of=mnist2tlhi and mnist2tllo];
\addplot[draw=none,forget plot,name path=mnist2vlhi] table[header=false,x index=0,y index=30]{figs_tex/data_mnist.dat};
\addplot[draw=none,forget plot,name path=mnist2vllo] table[header=false,x index=0,y index=29]{figs_tex/data_mnist.dat};
\addplot[tabgreen,opacity=0.15,forget plot] fill between[of=mnist2vlhi and mnist2vllo];
\addplot[tabblue,solid,thick,forget plot] table[header=false,x index=0,y index=1]{figs_tex/data_mnist.dat};
\addplot[tabblue,dashed,thick,forget plot] table[header=false,x index=0,y index=4]{figs_tex/data_mnist.dat};
\addplot[taborange,solid,thick,forget plot] table[header=false,x index=0,y index=13]{figs_tex/data_mnist.dat};
\addplot[taborange,dashed,thick,forget plot] table[header=false,x index=0,y index=16]{figs_tex/data_mnist.dat};
\addplot[tabgreen,solid,thick,forget plot] table[header=false,x index=0,y index=25]{figs_tex/data_mnist.dat};
\addplot[tabgreen,dashed,thick,forget plot] table[header=false,x index=0,y index=28]{figs_tex/data_mnist.dat};
\nextgroupplot[ylabel={Error (1 - Accuracy)}]
\addplot[draw=none,forget plot,name path=mnist0tehi] table[header=false,x index=0,y index=9]{figs_tex/data_mnist.dat};
\addplot[draw=none,forget plot,name path=mnist0telo] table[header=false,x index=0,y index=8]{figs_tex/data_mnist.dat};
\addplot[tabblue,opacity=0.15,forget plot] fill between[of=mnist0tehi and mnist0telo];
\addplot[draw=none,forget plot,name path=mnist0vehi] table[header=false,x index=0,y index=12]{figs_tex/data_mnist.dat};
\addplot[draw=none,forget plot,name path=mnist0velo] table[header=false,x index=0,y index=11]{figs_tex/data_mnist.dat};
\addplot[tabblue,opacity=0.15,forget plot] fill between[of=mnist0vehi and mnist0velo];
\addplot[draw=none,forget plot,name path=mnist1tehi] table[header=false,x index=0,y index=21]{figs_tex/data_mnist.dat};
\addplot[draw=none,forget plot,name path=mnist1telo] table[header=false,x index=0,y index=20]{figs_tex/data_mnist.dat};
\addplot[taborange,opacity=0.15,forget plot] fill between[of=mnist1tehi and mnist1telo];
\addplot[draw=none,forget plot,name path=mnist1vehi] table[header=false,x index=0,y index=24]{figs_tex/data_mnist.dat};
\addplot[draw=none,forget plot,name path=mnist1velo] table[header=false,x index=0,y index=23]{figs_tex/data_mnist.dat};
\addplot[taborange,opacity=0.15,forget plot] fill between[of=mnist1vehi and mnist1velo];
\addplot[draw=none,forget plot,name path=mnist2tehi] table[header=false,x index=0,y index=33]{figs_tex/data_mnist.dat};
\addplot[draw=none,forget plot,name path=mnist2telo] table[header=false,x index=0,y index=32]{figs_tex/data_mnist.dat};
\addplot[tabgreen,opacity=0.15,forget plot] fill between[of=mnist2tehi and mnist2telo];
\addplot[draw=none,forget plot,name path=mnist2vehi] table[header=false,x index=0,y index=36]{figs_tex/data_mnist.dat};
\addplot[draw=none,forget plot,name path=mnist2velo] table[header=false,x index=0,y index=35]{figs_tex/data_mnist.dat};
\addplot[tabgreen,opacity=0.15,forget plot] fill between[of=mnist2vehi and mnist2velo];
\addplot[tabblue,solid,thick,forget plot] table[header=false,x index=0,y index=7]{figs_tex/data_mnist.dat};
\addplot[tabblue,dashed,thick,forget plot] table[header=false,x index=0,y index=10]{figs_tex/data_mnist.dat};
\addplot[taborange,solid,thick,forget plot] table[header=false,x index=0,y index=19]{figs_tex/data_mnist.dat};
\addplot[taborange,dashed,thick,forget plot] table[header=false,x index=0,y index=22]{figs_tex/data_mnist.dat};
\addplot[tabgreen,solid,thick,forget plot] table[header=false,x index=0,y index=31]{figs_tex/data_mnist.dat};
\addplot[tabgreen,dashed,thick,forget plot] table[header=false,x index=0,y index=34]{figs_tex/data_mnist.dat};
\end{groupplot}
% Shared legend, both rows centred on the two-panel midline (row 1: three colours; row 2: line styles).
\path (group c1r1.north west) -- (group c2r1.north east) coordinate[midway] (legmnistmid);
\node[anchor=south, font=\small] at ([yshift=4mm]legmnistmid) {%
  \begin{tabular}{@{}c@{}}
    \tikz\draw[tabblue,thick](0,0)--(0.45,0);~SBQE\hspace{1.2em}%
    \tikz\draw[taborange,thick](0,0)--(0.45,0);~Ampl. encoding\hspace{1.2em}%
    \tikz\draw[tabgreen,thick](0,0)--(0.45,0);~MLP\\[2pt]
    \tikz\draw[black,thick](0,0)--(0.45,0);~train\hspace{1.2em}%
    \tikz\draw[black,thick,dashed](0,0)--(0.45,0);~validation
  \end{tabular}};
\end{tikzpicture}%
}
  \caption{Learning curves on the MNIST benchmark ($8$ qubits, $4$ layers).
  Left panel: cross-entropy loss; right panel: classification error ($1-$accuracy).
  Within each panel, solid lines are training and dashed lines validation, both averaged over 10 runs; shaded bands show one standard deviation, and colors distinguish the three models as in the legend.}
  \label{fig:mnist_curves}
\end{figure*}

\paragraph*{MNIST.}
On the standard handwritten-digit benchmark, SBQE reaches
\(90.25\%\pm0.18\%\) accuracy, again the highest of the three models,
exceeding amplitude encoding \(88.17\%\pm0.09\%\) by 2.1 percentage points
and the width-matched classical network \(89.95\%\pm0.14\%\) by 0.3.
Convergence curves are shown in Figure~\ref{fig:mnist_curves}.
Evaluated under the identical protocol used for the other datasets, this
confirms that the advantage of shot-based mixing over amplitude encoding is
not specific to the smaller Semeion and Fashion suites.

\paragraph*{Statistical comparison.}
Two-tailed paired \(t\)-tests over the 10 seeds confirm that SBQE is
significantly better than amplitude encoding on all three datasets
(\(p<0.01\) for Semeion and Fashion-MNIST, \(p\approx2\times10^{-10}\) for
MNIST); against the width-matched linear network the difference is not
significant for Semeion but is for Fashion-MNIST (\(p<0.05\)) and MNIST
(\(p\approx2\times10^{-3}\)).
Taken together, the results support our central hypothesis that shot
allocation is a competitive information channel for near-term quantum
classifiers.

\subsection{Contribution of the quantum layer}\label{sec:quantum_layer}

A natural question is whether the advantage of SBQE stems from the quantum
circuit itself or merely from the trainable classical layer that maps the
reduced features to $\vect{p}(\vect{x})$.  To isolate the circuit's
contribution, we replace it with a same-shape classical map of
$\vect{p}(\vect{x})$, a trainable linear layer
$2^{n_{\mathrm{q}}}\!\to\!n_{\mathrm{q}}$, and sweep its rank, holding the
SBQE front-end and readout fixed.  As shown in
Figure~\ref{fig:rank_sweep}, the quantum layer, whose mixing stage carries
only $3(n_{\mathrm{layers}}+1)\,n_{\mathrm{q}}=120$ trainable angles,
matches or exceeds classical maps that use an order of magnitude more
parameters; at the quantum layer's own parameter budget the classical
stand-in reaches only \(57\%\)--\(67\%\).  The quantum circuit, therefore,
supplies representational capacity beyond that of the classical front-end
alone, confirming that the separation is driven by the circuit rather than
by the preprocessing.

\begin{figure}[t]
  \centering
  % Auto-generated (pgfplots): Figure 5, rank sweep. Single-column (\columnwidth).
% Preamble requirements are listed in figs_tex/README_pgfplots.txt.
% Requires \bigstar (amssymb) and \scalebox (graphicx); both are in paper_revised.tex.
\resizebox{\columnwidth}{!}{%
\begin{tikzpicture}
\begin{axis}[
    width=9cm, height=5.9cm,
    xmode=log, log ticks with fixed point,
    x tick label style={/pgf/number format/1000 sep={}},
    xlabel={Mid-layer trainable parameters ($2^{n_{\mathrm{q}}}\!\to\!n_{\mathrm{q}}$ map)},
    ylabel={Test accuracy (\%)},
    xmin=95, xmax=2700, ymin=54, ymax=93,
    ytick={55,60,65,70,75,80,85,90},
    grid=both, grid style={gray!25},
    tick align=outside, tick pos=left,
    legend cell align=left, legend columns=2,
    legend style={draw=none, fill=none, font=\scriptsize,
        at={(0.5,-0.32)}, anchor=north, /tikz/every even column/.append style={column sep=10pt}},
    label style={font=\small}, tick label style={font=\footnotesize},
    error bars/error bar style={line width=0.6pt}, error bars/error mark options={mark size=2pt},
]
% Reference line at the quantum layer's parameter budget (matches the PNG).
\addplot[gray, densely dotted, thick, forget plot, mark=none] coordinates {(120,54) (120,93)};
% Fashion: classical low-rank (line + circle markers)
\addplot[taborange, thick, mark=*, mark size=2.2pt,
    error bars/.cd, y dir=both, y explicit]
  coordinates {(272,66.82) +- (0,2.17) (536,78.25) +- (0,0.40) (1064,80.72) +- (0,0.24) (2056,81.25) +- (0,0.22)};
\addlegendentry{Fashion: classical low-rank}
% Fashion: quantum layer (large filled star)
\addplot[taborange, only marks, mark=text, text mark={\textcolor{taborange}{\scalebox{1.7}{$\bigstar$}}},
    error bars/.cd, y dir=both, y explicit]
  coordinates {(120,80.95) +- (0,0.10)};
\addlegendentry{Fashion: quantum layer (SBQE)}

% MNIST: classical low-rank (line + circle markers)
\addplot[tabblue, thick, mark=*, mark size=2.2pt,
    error bars/.cd, y dir=both, y explicit]
  coordinates {(272,56.86) +- (0,2.13) (536,83.07) +- (0,0.60) (1064,89.02) +- (0,0.30) (2056,90.64) +- (0,0.15)};
\addlegendentry{MNIST: classical low-rank}
% MNIST: quantum layer (large filled star)
\addplot[tabblue, only marks, mark=text, text mark={\textcolor{tabblue}{\scalebox{1.7}{$\bigstar$}}},
    error bars/.cd, y dir=both, y explicit]
  coordinates {(120,90.25) +- (0,0.18)};
\addlegendentry{MNIST: quantum layer (SBQE)}
\end{axis}
\end{tikzpicture}%
}
  \caption{Test accuracy of the SBQE quantum layer (stars) compared with same-shape classical low-rank maps of $\vect{p}(\vect{x})$ ($2^{n_{\mathrm{q}}}\!\to\!n_{\mathrm{q}}$) as a function of the mixing layer's trainable-parameter count ($n_{\mathrm{q}}=8$ qubits, 4 layers; 10 seeds, best-validation weights).  The quantum layer reaches its accuracy with far fewer parameters than a classical map of comparable performance, indicating that the circuit (not the front-end or the PCA reduction) carries the class separation.}
  \label{fig:rank_sweep}
\end{figure}

% ================================================================
\section{Discussion}\label{sec:discussion}

SBQE uses an overlooked resource, \emph{classically allocated shots}, to sidestep the data-loading bottleneck.
By eliminating encoding gates, we reduce both depth and coherent error accumulation, rendering the scheme compatible with today's $\sim\!100$-gate error budgets.
Further, SBQE meshes naturally with measurement-based and classical-shadow post-processing; extending the framework to error-mitigated estimators is a promising direction.  Noise-induced regularization techniques \cite{kuzmin2025method} could be applied on top of SBQE to further improve generalization on noisy hardware.  Likewise, information-plane analysis, which quantify how strongly a model compresses its input and feed this signal back into training \cite{haboury2024information}, offer a natural tool for monitoring and tuning the learned probability map $\vect{p}(\vect{x})$.

Another point concerns linearly separable data.  Standard variational quantum circuits with angle encoding produce truncated Fourier series and are known to struggle with linear decision boundaries \cite{schuld2021effect,kordzanganeh2023exponentially,bowles2024better,wiedmann2024fourieranalysisvariationalquantum}.  Because SBQE encodes data through classical probabilities rather than through rotation angles, the resulting model class is not restricted to periodic functions, and preliminary experiments on synthetic two-dimensional tasks confirm that SBQE can separate linear data with ease.

Beyond the image benchmarks presented here, SBQE is well suited to high-dimensional inputs.  Since each initial state in the pool corresponds to one component of the probability vector, a register of $q$ qubits with $2^{q}$ basis states can encode up to $2^{q}$ features without any encoding gates.  For instance, the 784 pixels of an MNIST image fit into a 10-qubit first layer ($784 < 2^{10}=1024$).  This logarithmic scaling of qubits with feature count is shared with amplitude encoding but achieved here without the associated circuit depth.

\emph{Limitations.}
(i) Shot budgets grow with feature dimension if one insists on low-variance multinomial sampling; at a minimum, the total shot count must be large enough that each datum produces a distinguishable distribution.  How expressivity degrades as the shot budget decreases is an important open question.  Stratified resampling strategies may help.
(ii) The preprocessing step requires choosing an appropriate probability mapping $\vect{p}(\vect{x})$ for each dataset.  Functions such as softmax introduce their own nonlinearity, which is useful given that the quantum layer itself is linear in the probabilities, but a more principled selection method would be valuable.
(iii) Hardware constraints on state-preparation variety $\{\ket{\psi_j}\}$ could limit expressivity; adaptive state pools deserve exploration.  Superposition of parameterized circuits \cite{patapovich2025superposed} is one route toward richer initial-state families without proportional depth overhead.
(iv) Multilayer SBQE requires multiple sequential passes through the quantum device, which can be costly in wall-clock time.

% ================================================================
\section{Conclusion}\label{sec:conclusion}

We have introduced SBQE, a data-loading method that replaces gate-heavy encodings with probabilistic shot allocations.
The resulting mixed-state formalism is structurally equivalent to a classical multilayer perceptron whose weight matrices live inside a quantum circuit, and it composes naturally with variational layers.
On three image-classification benchmarks, SBQE matches or outperforms both amplitude encoding and a width-matched classical network while requiring zero encoding gates and staying within NISQ depth budgets.
These results suggest that the shot degree of freedom, already present in every quantum experiment, is a practical channel for data encoding on current hardware.

% ================================================================
\section*{Funding}
The authors have nothing to report.

\section*{Conflicts of Interest}
The authors declare no conflicts of interest.

\section*{Data Availability Statement}
The data that support the findings of this study are available from the corresponding author upon reasonable request.

% ================================================================

\bibliographystyle{apsrev4-2}
\bibliography{ref}

\end{document}